\documentclass[aps,prb,reprint,longbibliography,superscriptaddress]{revtex4-2}
\usepackage{mathrsfs}
\usepackage{amsmath,gensymb}
\usepackage{amsfonts}
\usepackage{amssymb}
\usepackage{amsthm}
\usepackage{graphicx}
\usepackage{natbib}
\usepackage{color}
\usepackage{hyperref}
\usepackage{bm}
\usepackage[caption=false]{subfig}
\usepackage{verbatim}
\usepackage[normalem]{ulem}

\begin{document}
\title{Andreev bound states at nonmagnetic impurities in superconductor/antiferromagnet heterostructures}

\author{G. A. Bobkov}
\affiliation{Moscow Institute of Physics and Technology, Dolgoprudny, 141700 Moscow region, Russia}

\author{I.V. Bobkova}
\affiliation{Moscow Institute of Physics and Technology, Dolgoprudny, 141700 Moscow region, Russia}
\affiliation{National Research University Higher School of Economics, 101000 Moscow, Russia}

\author{A. M. Bobkov}
\affiliation{Moscow Institute of Physics and Technology, Dolgoprudny, 141700 Moscow region, Russia}


\begin{abstract}
Andreev bound states can occur at single impurities in superconductors if the impurities suppress superconductivity for a given system. In particular, well-known Yu-Shiba-Rusinov states occur at magnetic impurities in conventional $s$-wave superconductors. Here we demonstrate that nonmagnetic impurities in S/AF heterostructures with conventional intraband $s$-wave pairing also produce Andreev bound states. 
Analogously to the Yu-Shiba-Rusinov bound states the bound states in S/AF bilayers are spin split, but the spin of a particular bound state is determined by the sublattice to which the impurity belongs. The standard decay of the bound state LDOS is superimposed by atomic oscillations related to the staggered character of the exchange field in the host material and by another oscillating pattern produced by finite-momentum N\'eel triplet pairing generated at the impurity. 
\end{abstract}

\maketitle

\section {Introduction}

Impurities in superconductors are one of the
most important tools for identifying the nature of the
pairing state and microscopic properties of superconductors and for realization of topological superconductivity \cite{Balatsky2006,Yazdani1997,Hudson2001,Grothe2012,Nadj-Perge2014,Pawlak2016,Schneider2021}. A lot of efforts is devoted to investigation of how different types of impurities influence on the critical superconducting temperature for different types of superconducting pairing. In  conventional  $s$-wave superconductors non-magnetic impurities do not suppress superconductivity according to the Anderson’s theorem \cite{Anderson1959}. However, magnetic impurities are pair-breaking for conventional  $s$-wave superconductors and suppress their critical temperature \cite{Abrikosov1961,Balatsky2006}. For unconventional superconductors with anisotropic  \cite{Mineev_book,Balatsky2006} or $s$-wave odd-parity pairing \cite{Michaeli2012,Cavanagh2020,Cavanagh2021,Dentelski2020,Sato2020} even nonmagnetic impurities can be pair-breaking.     

The related important problem is study of Andreev bound states, which can occur in the vicinity of single impurities if the impurities suppress superconductivity for a given system. The single-impurity Andreev bound states have attracted much attention
over the last several decades \cite{Balatsky2006}. The magnetic impurities break the time-reversal symmetry and for this reason they are pair-breaking even for conventional $s$-wave superconductors. Well-known Yu-Shiba-Rusinov states occur at magnetic impurities \cite{Luh1965,Shiba1968,Rusinov1969}. On chains of magnetic impurities they can form topological bands due to overlapping of bound states at separate impurities \cite{Nadj-Perge2014,Pawlak2016,Schneider2021}. For $d$-wave superconductors even nonmagnetic impurities are pair-breakers and produce bound states due to the fact that the change of
the quasiparticle momentum upon scattering disrupts
the phase assignment for particular directions of the momenta in such a pairing state \cite{Balatsky1995,Balatsky2006}. In addition, the nonmagnetic impurities can result in the superconductivity suppression and impurity-induced Andreev bound states in multi-band superconductors \cite{Tsai2009,Wang2018,Zhu2023}. In unconventional superconductors they serve as a test of the pairing symmetry \cite{Hudson2001,Grothe2012}. 

\begin{figure}[tb]
	\begin{center}
		\includegraphics[width=85mm]{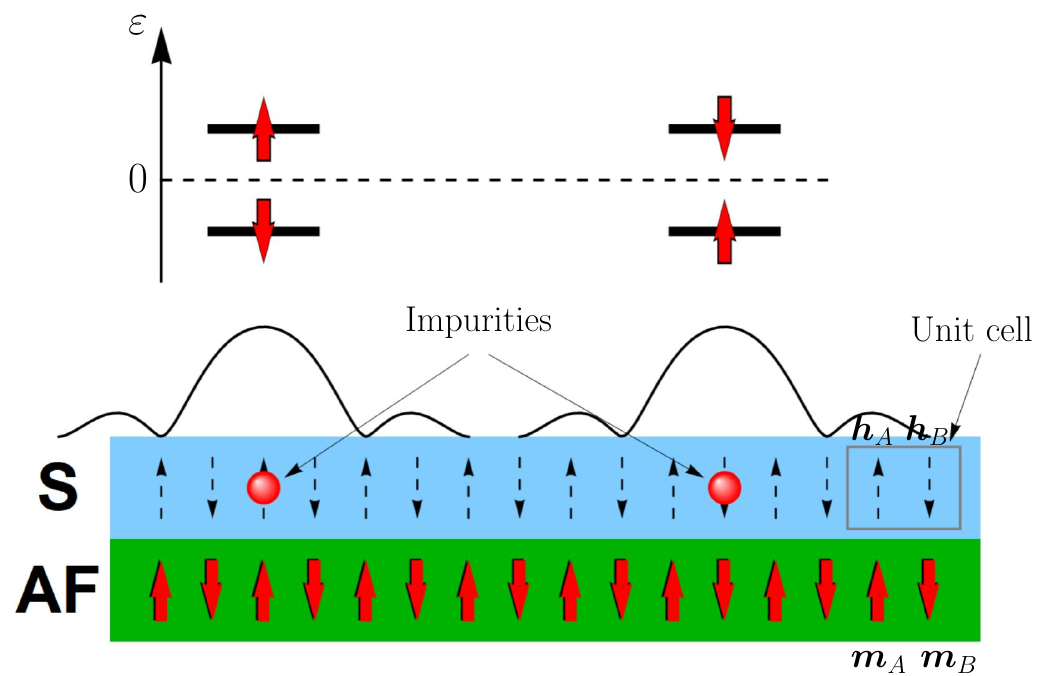}
\caption{Sketch of the system under consideration. Insulating two-sublattice antiferromagnet (AF) with staggered magnetization $\bm m_A = -\bm m_B$ induces a staggered exchange field $\bm h_A = -\bm h_B$ via the proximity effect in the adjacent thin superconductor (S). The unit cell is shown by a rectangular. An impurity can occupy site $A$ or $B$ in the S layer. The both possible variants are shown by red balls. The LDOS of Andreev bound states localized at the corresponding impurity is shown schematically. The energy spectrum of the bound states with the appropriate spin structure (red arrows) is also shown above the corresponding impurity.}
 \label{fig:sketch}
	\end{center}
\end{figure}

Here we demonstrate that impurity-induced Andreev bound states  at {\it nonmagnetic} impurities can also occur in superconductor/antiferromagnet (S/AF) heterostructures with conventional intraband $s$-wave pairing. The system is sketched in Fig.~\ref{fig:sketch} and represents a thin-film bilayer composed of a superconductor and a two-sublattice antiferromagnet. The general physical argument allowing for the bound state at a nonmagnetic impurity in such a system is the following. The ideal bilayer in the absence of impurities is symmetric under simultaneous action of time reversal and sublattice interchange. The presence of impurity breaks this symmetry. As a result, for conduction electrons the impurity can be viewed as effectively magnetic.  Analogously to the Yu-Shiba-Rusinov bound states the bound states in S/AF bilayers are spin split, but the spin of a particular bound state is determined by the sublattice to which the impurity belongs, see Fig.~\ref{fig:sketch} for illustration. 

The presence of Andreev bound states at single nonmagnetic impurities in S/AF bilayers is in agreement with the behavior of the superconducting critical temperature of such systems in the presence of random disorder, which has already been studied \cite{Buzdin1986,Bobkov2023,Fyhn2023}. In Ref.~\cite{Bobkov2023} it was shown that nonmagnetic impurities can suppress or {\it enhance} superconductivity depending on the value of the chemical potential. At $\mu \lesssim T_{c0}$, where $T_{c0}$ is a critical temperature of the superconductor, the nonmagnetic impurities enhance superconductivity of S/AF bilayers. It is connected with the presence of so-called N\'eel triplet correlations \cite{Bobkov2022_Neel}, which are destructive to singlet superconductivity in the S layer, but are actually interband correlations and therefore are suppressed by nonmagnetic disorder. On the contrary, if $\mu \gg T_{c0}$ the superconductivity is suppressed by  nonmagnetic disorder. Here we demonstrate that the same sensitivity to the value of the chemical potential occurs in the problem of a single impurity: the bound states only exist at $\mu \gg T_{c0}$, when the impurities are suppressing for superconductivity.

The paper is organized as follows. In Sec.~\ref{model} we describe the considered model and formulate the $T$-matrix approach generalized for theoretical description of single impurity problems in the two-sublattice formalism. In Sec.~\ref{energies} we discuss the dependence of bound state energies on all significant physical parameters and present dependencies of the impurity-induced local density of states (LDOS) on the quasiparticle energy. Sec.~\ref{spatial} is devoted to discussion of the spatial distribution of the LDOS around an impurity. Our conclusions are formulated in Sec.~\ref{conclusions}. The Appendix provides additional information  on the derivation of the two-sublattice $T$-matrix formalism, phase diagrams of the regions, where the bound state exist, the spin and spatial structure of the impurity-induced LDOS. 

\section{System and method}
\label{model}

In the considered thin-film S/AF bilayer, see Fig.~\ref{fig:sketch} the antiferromagnet is assumed to be an insulator. The magnetism is staggered and the S/AF interface is fully compensated, that is the interface magnetization has zero average value. The sites in the superconductor are marked by the radius-vector $\bm r = (x, y, z)^T$, where $x,y,z$ are integer numbers, the interface is in the $(x,y)$-plane. The influence of the antiferromagnetic insulator on the thin S layer with the thickness smaller than the superconducting coherence length $\xi$ can be described by the N\'eel-type exchange field  $\bm h_{\bm r}^N =  (-1)^{x+y+z} \bm h $ \cite{Bobkov2022_Neel}, what leads to the following effective lattice Hamiltonian of the S layer: 
\begin{align}
\hat H= &- t \sum \limits_{\langle \bm{r}\bm{r'}\rangle,\sigma} \hat c_{\bm{r} \sigma}^{\dagger} \hat c_{\bm{r'} \sigma} + \sum \limits_{\bm{r}} (\Delta \hat c_{\bm{r}\uparrow}^{\dagger} \hat c_{\bm{r}\downarrow}^{\dagger} + H.c.) - \mu \sum \limits_{\bm{r}, \sigma} \hat n_{\bm{r}\sigma}\nonumber \\ 
&-
 \sum \limits_{\bm{r},\alpha \beta} \hat c_{\bm{r}\alpha}^{\dagger} (\bm{h}_{\bm{r}}^N \bm{\sigma})_{\alpha \beta} \hat c_{\bm{r}\beta} + \sum \limits_{\sigma} U_0 c_{\bm{r_{imp}},\sigma}^{\dagger}c_{\bm{r_{imp}},\sigma} 
\label{ham}
\end{align}
where $\langle \bm r \bm r' \rangle $ means summation over the nearest neighbors, $\hat{c}_{\bm r \sigma}^{\dagger}(\hat{c}_{\bm r\sigma})$ is the creation (annihilation) operator for an electron with spin $\sigma$ at site $\bm r$. $t$ parameterizes the hopping between adjacent sites, $\Delta$ accounts for on-site s-wave pairing. Since the superconducting layer is thin with respect to  $\xi$, $\Delta$ and $\bm h$ are assumed to be homogeneous along the $z$-direction. $\mu$ is the electron chemical potential. $\hat n_{\bm r \sigma} = \hat c_{\bm r \sigma}^{\dagger} \hat c_{\bm r \sigma}$ is the particle number operator, $\bm \sigma = (\sigma_x, \sigma_y, \sigma_z)^T$ are Pauli matrices in spin space. The lattice constant is denoted by $a$. $U_0$ is a potential of the single nonmagnetic impurity, which is located at site $\bm r_{imp}$. 

To calculate the bound state energies and LDOS around a single impurity we generalized $T$-matrix formalism \cite{Balatsky2006} for taking into account the antiferromagnetic character of the host material. It is done on the basis of the Gor'kov Green's functions in two-sublattice framework\cite{Bobkov2022_Neel,Bobkov2023}. Relegating technical details of the derivations to the Appendix \ref{T-matrix}, here we present the resulting equations. In the framework of the two-sublattice formalism we choose the unit cell with two sites belonging to two sublattices $A$ and $B$, as shown in Fig.~\ref{fig:sketch}. The Green's function is a $8 \times 8$ matrix in the direct product of spin, particle-hole and sublattice spaces. Therefore, in addition to the Pauli matrices $\bm \sigma = (\sigma_x, \sigma_y, \sigma_z)^T$ in spin space we  define the Pauli matrices $\bm \tau = (\tau_x, \tau_y, \tau_z)^T$ in particle-hole space and $\rho = (\rho_x, \rho_y, \rho_z)^T$ in sublattices space. 

In the framework of the $T$-matrix approach the retarded Green's function takes the form:
\begin{align}
    \check G_{\bm i \bm i}=\check G_{\bm i \bm i}^0+\check G_{\bm i \bm i_{imp}}^0 \check T\check G_{\bm i_{imp} \bm i}^0,
    \label{G_T_matrix}
\end{align}
where $\bm i \equiv \bm r_A$ is the coordinate of a unit cell coinciding with the radius-vector of its A-site. $\check G_{\bm i \bm j}^0$ is the homogeneous Green's function  of the S/AF bilayer in the absence of the impurity \cite{suppl} and the $T$-matrix takes the form
\begin{align}
    \check T=(1-\check U_0 \check G_{\bm i_{imp} \bm i_{imp}}^0 )^{-1}\check U_0
\end{align}
with $\check U_0$ being a matrix of impurity potential, which takes the form:
\begin{align}
    \check U_0 =U_0 \frac{(\rho_x\pm i\rho_y)}{2}. 
    \label{impurity_potential}
\end{align}
The sign $\pm$ describes impurity located at a site $A(B)$. For definiteness further we assume that the impurity is located at $A$-site. The results for impurities located at a site $B$ are obtained by $\bm h \to - \bm h$. Then the bound state energies are determined by the following equation:
\begin{align}
    \det (1-\check U_0  \check G_{\bm i_{imp} \bm i_{imp}}^0 )=0,
    \label{energy_bound_state}
\end{align}
$\check G_{\bm i_{imp} \bm i_{imp}}^0$  can be found analytically, see Appendix \ref{T-matrix}, it is diagonal in spin space $\check G_{\bm i_{imp} \bm i_{imp}}^0 = \check G_{\bm i_{imp} \bm i_{imp},\uparrow}^0 (1+\sigma_z)/2 + \check G_{\bm i_{imp} \bm i_{imp},\downarrow}^0 (1-\sigma_z)/2 $ and its spin-up and spin-down components take the form:
\begin{align}
    \check G_{\bm i_{imp} \bm i_{imp},\sigma}^0=G_{0x}^\sigma \tau_0\rho_x+G_{yx}^\sigma \tau_y\rho_x+G_{zx}^\sigma \tau_z\rho_x+  \\ \nonumber
+G_{0y}^\sigma \tau_0\rho_y+G_{yy}^\sigma\tau_y\rho_y+G_{zy}^\sigma\tau_z\rho_y
    \label{GF_homogeneous}
\end{align}

\begin{align}
    \begin{cases}
        G_{0x}^\sigma=\mu((-h^2+\Delta^2-\varepsilon^2+\mu^2)I_1-I_2) \\
        G_{yx}^\sigma=-i\Delta((-h^2+\Delta^2-\varepsilon^2+\mu^2)I_1+I_2) \\
        G_{zx}^\sigma=\varepsilon((-h^2-\Delta^2+\varepsilon^2-\mu^2)I_1-I_2) \\
        G_{0y}^\sigma=2ih \sigma \varepsilon \mu I_1 \\
        G_{yy}^\sigma=2h \sigma\Delta \varepsilon I_1 \\
        G_{zy}^\sigma=i h \sigma ((h^2-\Delta^2-\varepsilon^2-\mu^2)I_1+I_2),
                    \end{cases}
            \end{align}
where $\varepsilon$ is a quasiparticle energy, $\sigma = \pm 1$ as a factor in the expression for spin-up(down)-components, $h = |\bm h|$, $I_1=-\frac{2\sqrt{2}i\pi}{(\alpha_1+\alpha_2)\alpha_1\alpha_2}$, $I_2=\frac{\sqrt{2}i\pi}{\alpha_1+\alpha_2}$, $\alpha_{1,2}=\sqrt{-\beta_1\mp\sqrt{\beta_1^2-4\beta_0}}$, $\beta_0=h^4+(\Delta^2-\varepsilon^2+\mu^2)^2-2h^2(\Delta^2+\varepsilon^2+\mu^2)$, $\beta_1=2(\Delta^2-\varepsilon^2-\mu^2+h^2)$.

The superconducting order parameter is taken $\Delta = 0.1 t$ throughout the paper. In principle, it is also disturbed by the pair-breaking impurity and should be calculated self-consistently. But we do not perform the sel-consistency procedure, what is a quite standard approximation upon treating single-impurity problems because the suppression of the order parameter is
determined by the Fermi wavelength and does not affect
the position of the bound state \cite{Balatsky2006}.

\section{Bound state energies and energy-resolved DOS}

\label{energies}

\begin{figure}[tb]
	\begin{center}
		\includegraphics[width=85mm]{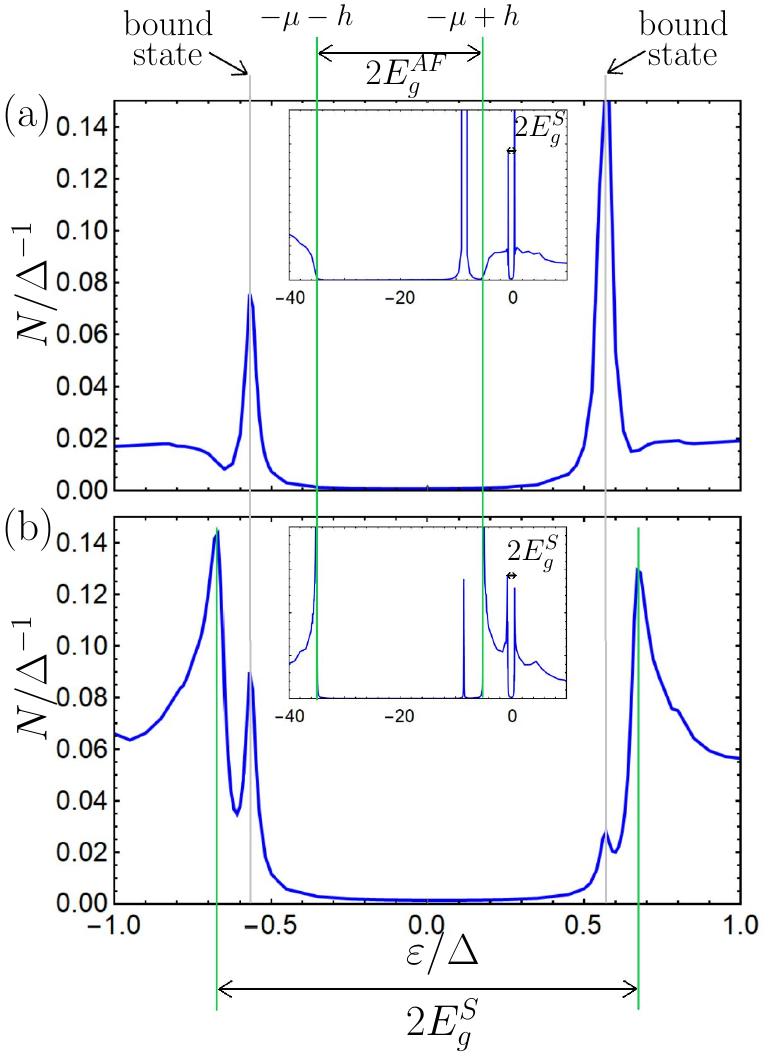}
\caption{LDOS as a function of energy. (a) LDOS at the impurity site. (b) LDOS at the nearest neighbor of the impurity. The inserts show the same LDOS for larger energy range. $2E_g^S$ is a superconducting gap at the Fermi level. The antiferromagnetic gap $2E_g^{AF}$ at $\varepsilon \in [-\mu - h, -\mu+h]$ and the non-superconducting additional bound state inside this gap are also seen. $t=10 \Delta$, $\mu=20\Delta$, $h=15\Delta$, $U_0 = 10\Delta$. The Dynes parameter, describing the level broadening, $\Gamma=0.02\Delta$, see Appendix \ref{T-matrix} for description.}
 \label{fig:LDOS_energy}
	\end{center}
\end{figure}

The LDOS at A and B sublattices can be calculated as:
\begin{align}
    N^{A,B}(\varepsilon,\bm i)= 
    -\frac{1}{\pi}{\rm Im}\left[{\rm Tr} [\frac{\check G_{\bm i \bm i} (\tau_0+\tau_z)(\rho_x \pm i\rho_y)}{4}]\right],
    \label{LDOS}
\end{align} 
The LDOS calculated at the impurity site and at the nearest neighbor of the impurity, is plotted in Figs.~\ref{fig:LDOS_energy}(a) and (b), respectively. The bound states are represented by the peaks at $\varepsilon = \pm \varepsilon_b$  with asymmetric heights  inside the superconducting gap. Higher outer peaks in panel (b) correspond to the superconducting gap $2E_g^S$. At the impurity site they are fully destroyed by the bound state [see panel (a)]. The asymmetry of the bound state peaks is connected with the overall particle-hole asymmetry of the LDOS. This asymmetry is clearly seen in the inserts, which show the LDOS for larger energy range. Here we can also see the antiferromagnetic gap $2E_g^{AF}$ at energies $\varepsilon \in [-\mu - h, -\mu+h]$ and an additional peak inside this gap. This peak corresponds to another impurity-induced bound state, which is not related to superconductivity and survives also in the normal state. We do not focus on this non-superconducting bound state here. 

\begin{figure}[tb]
	\begin{center}
		\includegraphics[width=85mm]{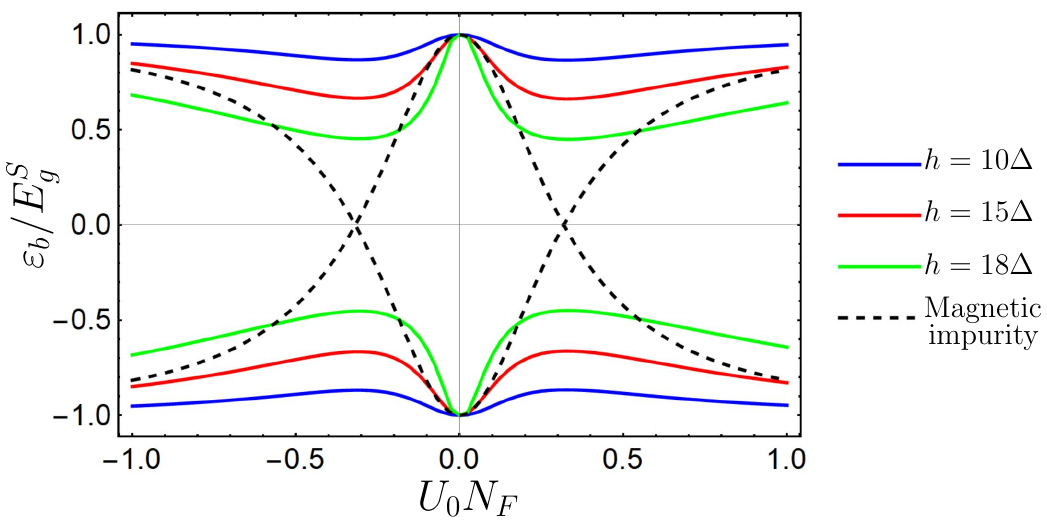}
\caption{Bound state energies as functions of the impurity strength. $\mu=20\Delta$. Different colors correspond to different $h$. Dashed lines represent bound state energies at a magnetic impurity with the same strength in a conventional $s$-wave superconductor.}
 \label{fig:energies}
	\end{center}
\end{figure}

As it was briefly mentioned in the introduction, the physical reason for appearance of these bound states is that the impurity in the S/AF host behaves like a magnetic impurity. Below we provide more details about this physical explanation. Energies of the bound states as functions of the impurity strength are plotted in Fig.~\ref{fig:energies}. It is seen that at stronger staggered effective exchange field $h$ the bound state is shifted deeper inside the superconducting gap region. For the chosen set of parameters we cannot consider $h>0.9\mu=18 \Delta$ because of the overall suppression of superconductivity by the N\'eel triplet correlations \cite{Bobkov2023}. For comparison the energies of the Yu-Shiba-Rusinov bound states at a magnetic impurity with the same strength in a conventional $s$-wave superconducting host are plotted by the dashed lines. Unlike the case of magnetic impurity our nonmagnetic impurities in S/AF bilayers are not able to support low and zero-energy bound states. In this sense one can say that they are weaker pair-breakers as compared to the magnetic impurities.

\begin{figure}[tb]
	\begin{center}
		\includegraphics[width=75mm]{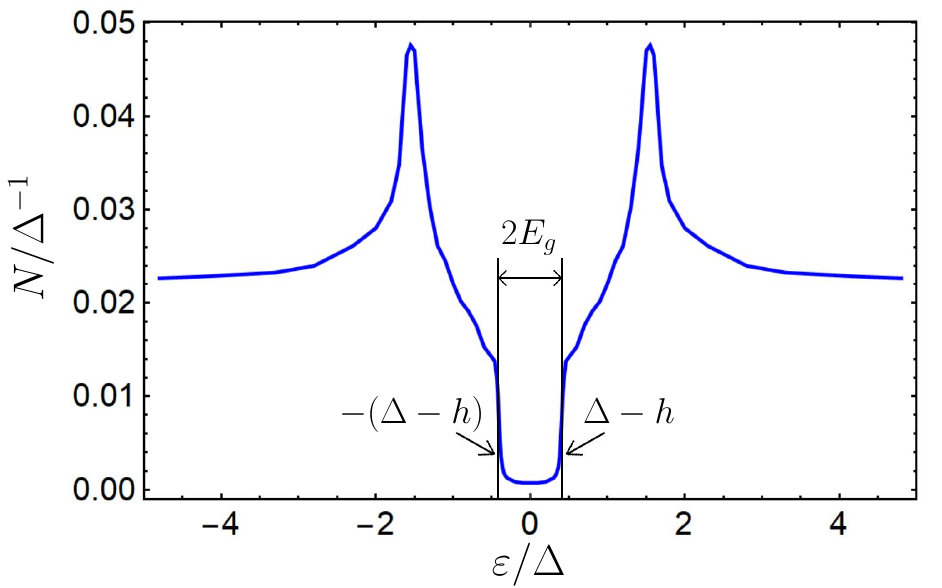}
\caption{LDOS as a function of energy.  At $\mu = 0$ there is the only gap $2E_g = 2(\Delta - h)$. In this case both antiferromagnetic and superconducting gaps are open at the Fermi level and, therefore, $2E_g$ is of mixed origin. $\mu=0$, $h=0.6\Delta$, $U_0 = 10\Delta$.}
 \label{fig:LDOS_energy2}
	\end{center}
\end{figure}

All the results discussed above are related to the case of strong chemical potential $\mu = 20\Delta$, which is far from the half-filling. For $\mu=0$ the LDOS $N^A(\varepsilon,\bm i_{imp})$ as a function of $\varepsilon$ is shown in Fig.~\ref{fig:LDOS_energy2}. One can see that if the chemical potential is close to half-filling $\mu=0$, the bound states do not appear. Therefore, one can conclude that whether or not an impurity in the S/AF bilayer becomes effectively magnetic depends on the value of the chemical potential. To understand the underlying physical reason let us consider the wave functions of Bloch electrons in a homogeneous superconductor in the presence of the staggered exchange field. They take the form:
\begin{eqnarray}
\left( \begin{array}{c} \hat \psi_{\bm i \sigma}^A \\ \hat \psi_{\bm i \sigma}^B
\end{array}
\right)(\bm p) =  \left( \begin{array}{c} \sqrt{1+\sigma h/(\mu+\varepsilon)} \\ \sqrt{1-\sigma h/(\mu+\varepsilon)}e^{-ip_y a_y}
\end{array}
\right) e^{i \bm p \bm i}, 
\label{eq:eigen_zero_SOC}
\end{eqnarray}
where it is assumed that the unit cell is chosen along the $y$-axis, $p_y$ and $a_y$ are the electron momentum component and the lattice constant along this axis, respectively. We are mainly interested in the low energies $\varepsilon \sim \Delta$. If $\mu \gg \Delta$ the probability density of spin-up electrons is presumably concentrated at sites $A$ and the probability density of spin down electrons is concentrated at sites $B$. Consequently, if the impurity is located at a site $A$, it interacts more strongly with spin-up electrons and vice versa \cite{Fyhn2023}. This is the reason of the effectively magnetic behavior of nonmagnetic impurities in S/AF bilayers. On the contrary, if $\mu=0$, then at $\varepsilon<0$ spin-up electrons are concentrated at $A$-sites, but at $\varepsilon>0$ spin-down electrons are concentrated there. Since all energies $\varepsilon \sim \Delta$ around the Fermi level contribute to superconducting pairing symmetrically, there is no difference in the spin-up and spin-down contribution to the interaction between the impurity and electrons. Therefore, impurities do not behave like effectively magnetic. 

\begin{figure}[tb]
	\begin{center}
		\includegraphics[width=75mm]{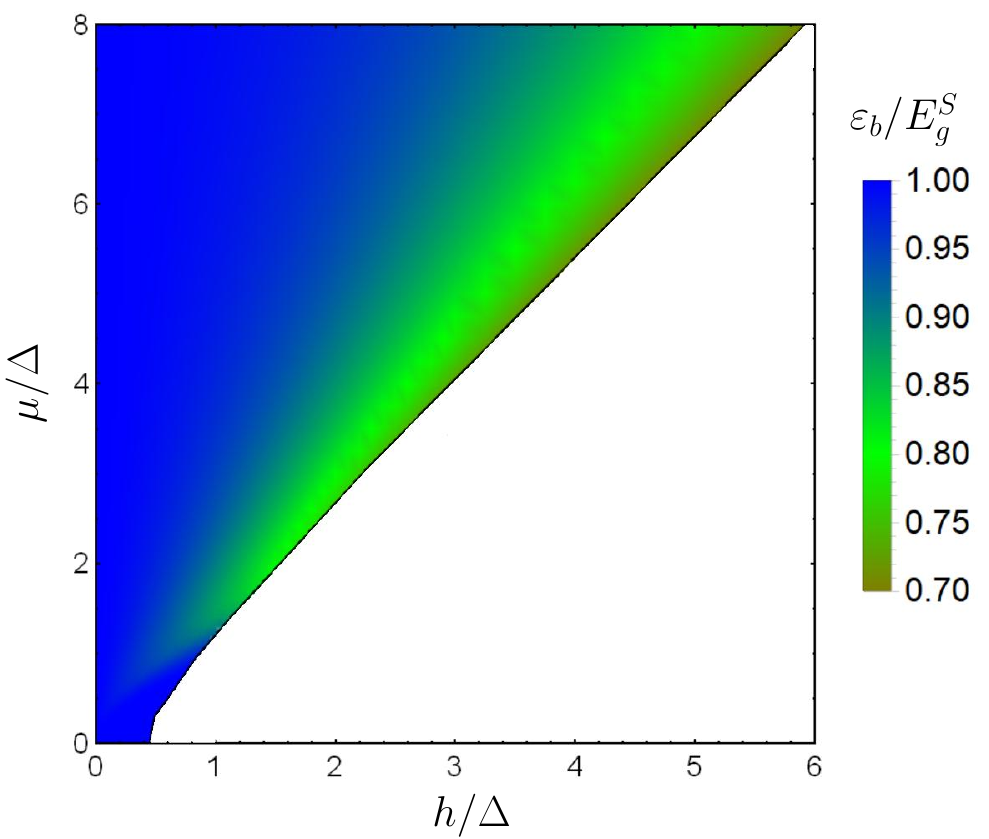}
\caption{Bound state energy $\varepsilon_b$ in the plane $(h,\mu)$. White region corresponds to fully suppressed superconductivity. $U_0N_F = 0.3$.}
 \label{fig:phase_diagram}
	\end{center}
\end{figure}

More quantitatively the regions of existence of the bound states and the dependence of the bound state energies on $\mu$ and $h$ can be summarized in the form of the phase diagram presented in Fig.~\ref{fig:phase_diagram}. In this Figure the bound state energy $\varepsilon_b$ is shown in the plane $(h,\mu)$. The white region corresponds to full suppression of  superconductivity by the N\'eel exchange field of the AF layer. Although in the present paper the superconducting order parameter is not calculated self-consistently, the boundary of the superconductivity suppression in the $(h,\mu)$-plane has already been calculated earlier \cite{Bobkov2023}. From Fig.~\ref{fig:phase_diagram} it is seen that at small $\mu \lesssim \Delta$ the bound state indeed merges with the superconducting gap $E_g^S$ for an arbitraty $h$. For larger $\mu$ the bound state appears. Its energy is governed by the parameter $h/\mu$ because it is this parameter, which is responsible for making the impurity effectively magnetic, see Eq.~(\ref{eq:eigen_zero_SOC}). The impurity strength $U_0$ is chosen to provide  maximal deviation of the bound state energy from the edge of the gap. In order to prove that the regions in the $(h,\mu)$-plane, where the bound states exist, do not depend qualitatively on the particular choice of $U_0$, and to provide more information on the phase diagram, we also plotted the curvature $\bigl|d^2\varepsilon_b/dU_0^2 |_{U_0=0}\bigr|$ in the plane $(h,\mu)$, see Appendix \ref{phase_diagram}. If this quantity is above zero, it indicates the existence of the bound states for a given set $(h, \mu)$ at an arbitrary value of $U_0$. 

\section{Spatial structure of the LDOS around impurity}

\label{spatial}

\begin{figure}[tb]
	\begin{center}
		\includegraphics[width=90mm]{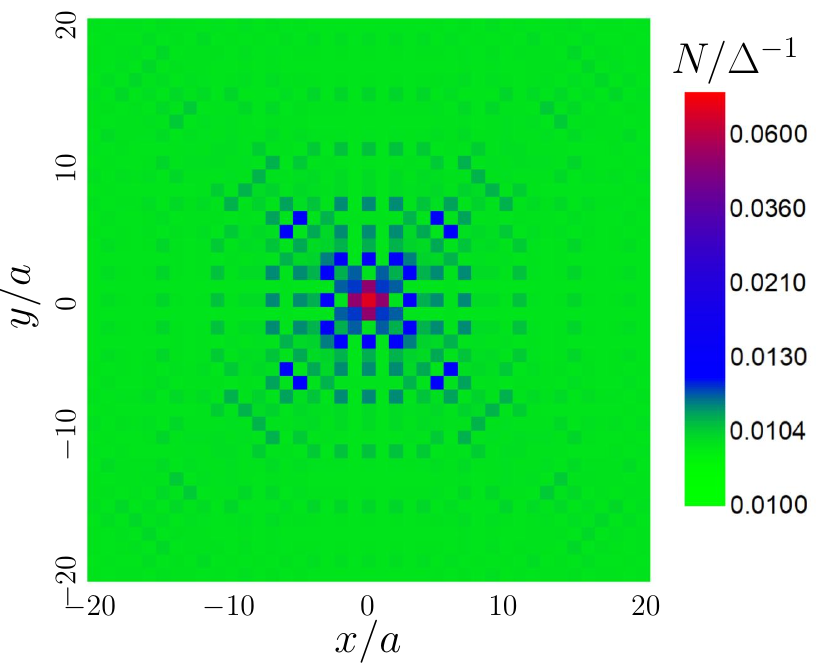}
\caption{LDOS at $\varepsilon = -\varepsilon_b$ as a function of coordinates. The impurity is at $A$-site of the unit cell $\bm i_{imp} = (0,0)$. $\mu=20\Delta$, $h=15\Delta$, $U_0 = 10\Delta$. }
 \label{fig:LDOS_spatial}
	\end{center}
\end{figure}

Spatial structure of the LDOS around impurity at the energy of the bound state $\varepsilon = -\varepsilon_b$ is calculated according to Eq.~(\ref{LDOS}) and is presented in Fig.~\ref{fig:LDOS_spatial}. The spatial region occupied by the bound state has a spatial scale of the order of the superconducting coherence length $\xi \sim v_F/\Delta \sim 2at/\Delta$, where $v_F \sim 2at$ is the Fermi velocity of an electron in the normal state of the superconductor. The exponential decay is superimposed by a power-law suppression analogously to the case of magnetic impurities in conventional superconductors \cite{Balatsky2006}. However, unlike the magnetic impurities in conventional superconductors here the LDOS has a "staggered" component, which oscillates between the sublattices. It is seen, that if the impurity is localized at $A$-site, the bound state LDOS is mainly concentrated at the $B$-sublattice everywhere except for the atomic-scale region near the impurity site. This fact is closely connected with the spin polarization of the bound state LDOS. The spin-resolved impurity-induced LDOS is presented in Fig.~\ref{fig:spin} in Appendix \ref{spin_structure}. It is seen that the bound states are indeed spin-split and if the impurity is at $A$-site the lower bound state $\varepsilon = -\varepsilon_b$ corresponds to the spin-down polarization (if the impurity is at $B$-site, the spin polarization of the bound states is reversed), as it is illustrated in Fig.~\ref{fig:sketch}. According to Eq.~(\ref{ham}) it is energetically favorable to concentrate the DOS for spin-down electrons at the B-sublattice to minimize the exchange energy. 

Another interesting feature of the spatial structure of the bound state LDOS is that the overall decay of the LDOS and "staggered" oscillations associated with the sublattice structure are also superimposed by  oscillations of a larger spatial scale compared to the atomic one, which is nevertheless significantly smaller than the superconducting coherence length scale. We associate these oscillations with the generation of finite-momentum N\'eel-type triplet correlations, which were originally predicted for S/AF bilayers with metallic antiferromagnets due to the Umklapp electron scattering processes at the AF/S interface \cite{Bobkov2023_2}. The period of oscillations is $L_{osc} = \pi v_F/ \sqrt{\mu^2 - h^2}$. Data presented in Fig.~\ref{fig:LDOS_spatial} are calculated at $h=1.5 t$ and $\mu = 2 t$. Then $L_{osc} \approx 4 a$, which is in agreement with the additional oscillation period seen in the figure. More detailed data proving that
the reason for the appearance of the additional period
of the LDOS oscillations is the finite-momentum N\'eel triplet pairing can be found in Appendix \ref{appendix_spatial}. These data include the Fourier transform of Fig.~\ref{fig:LDOS_spatial}, explicit spatial structure of the N\'eel triplets demonstrating the same period. We also extracted the corresponding period for different $(h,\mu)$ points and checked that they are in excellent agreement with the analytical expression for $L_{osc}$, see Appendix \ref{appendix_spatial}.  

\section {Conclusions}

\label{conclusions}

Basing on the $T$-matrix approach we demonstrated that nonmagnetic impurities in S/AF heterostructures with conventional intraband $s$-wave pairing produce Andreev bound states. The physical reason is that the nonmagnetic impurities in the presence of a staggered exchange field in the host material can behave like effectively magnetic due to the atomically oscillating, "staggered", character of wave functions of conduction electron. Whether or not an impurity in the S/AF bilayer becomes effectively magnetic depends on the value of the chemical potential. The bound states only exist at $\mu \gg T_{c0}$. It is in agreement with the behavior of the superconducting critical temperature of such systems in the presence of random disorder because it was shown earlier that only in this regime the nonmagnetic impurities are pair-breaking for conventional $s$-wave superconductivity. 

Analogously to the Yu-Shiba-Rusinov bound states the bound states in S/AF bilayers are spin split, but the spin of a particular bound state is determined by the sublattice to which the impurity belongs. The spatial structure of the bound state LDOS is also investigated. It is shown that the standard decay of the bound state LDOS is superimposed by atomic oscillations related to the staggered character of the exchange field in the host material and by another oscillating pattern produced by finite-momentum N\'eel triplet pairing generated at the impurity.

\begin{acknowledgments}
We are grateful to A.A. Golubov for many fruitful discussions. G.A.B and I.V.B. acknowledge the financial support from the Foundation for the Advancement of Theoretical Physics and Mathematics “BASIS” via the project 23-1-1-51-1.  The numerical calculations were supported by the Russian Science Foundation via the RSF project No.23-72-30004.  
\end{acknowledgments}

\appendix
\section{$T$-matrix approach for two-sublattice Green's functions}
\label{T-matrix}

Calculations of the bounds states and the LDOS have been performed in the framework of the $T$-matrix approach generalized for the antiferromagnetic host material. The general formalism of the Gor'kov Green's functions in two-sublattice framework have been developed earlier \cite{Bobkov2022_Neel,Bobkov2023}. The unit cell with two sites in it is chosen as shown in Fig.~1 of the main text. Introducing the two-sublattice Nambu spinor $\check c_{\bm i} = (\hat c_{{\bm i},\uparrow}^A, \hat c_{\bm i,\downarrow}^A, \hat c_{\bm i,\uparrow}^B,\hat c_{\bm i,\downarrow}^B, \hat c_{\bm i,\uparrow}^{A\dagger}, \hat c_{\bm i,\downarrow}^{A\dagger}, \hat c_{\bm i,\uparrow}^{B\dagger}, \hat c_{\bm i,\downarrow}^{B\dagger})^T$ and the Pauli matrices $\bm \sigma = (\sigma_x, \sigma_y, \sigma_z)^T$ in spin space, $\bm \tau = (\tau_x, \tau_y, \tau_z)^T$ in particle-hole space and $\rho = (\rho_x, \rho_y, \rho_z)^T$ in sublattices space, we define the retarded Green's function as 
\begin{eqnarray}
\check {G}_{\bm i \bm j}(t_1-t_2) = 
-\left(
\begin{array}{cc}
1 & 0 \\
0 & -i\sigma_y
\end{array}
\right)_\tau \rho_x \Theta(t_1-t_2) \times \nonumber \\ 
\left \langle \left\{ \check c_{\bm i}(t_1), \check c_{\bm j}^\dagger(t_2) \right\} \right \rangle 
\left(
\begin{array}{cc}
1 & 0 \\
0 & -i\sigma_y
\end{array}
\right)_\tau ,
\label{unitary}
\end{eqnarray}
where $\bm i$ is now the radius-vector of the full unit cell,  subscript $\tau$ means that the explicit matrix structure corresponds to the particle-hole space.
The Green's function is a $8 \times 8$ matrix in the direct product of spin, particle-hole and sublattice spaces.  We choose the $z$-axis along $\bm h$. The Fourier-transformed  Green's function $\check {G}_{\bm i \bm j}(\varepsilon) = \int e^{i \varepsilon (t_1-t_2)} \check {G}_{\bm i \bm j}(t_1-t_2) d(t_1-t_2)$ obeys the following equation: 
\begin{align}
    (\check H^0_{\bm i}-\check U_{\bm i})\check G_{\bm i \bm j}=\delta_{\bm i \bm j},
    \label{Green_general}
\end{align}
where $\check H_{\bm i}^0$ is the hamiltonian of the homogeneous S/AF bilayer, which acts on the Green's function as follows:
\begin{align}
    \check H_{\bm i}^0 \check G_{\bm i \bm j}=[(\varepsilon+i \Gamma)\tau_z+\mu+h\sigma_z\rho_z+\Delta i\tau_y]\rho_x \check G_{\bm i\bm j} + \check K_{\bm i} \check G_{\bm i \bm j} 
\end{align}
where $\Gamma>0$ is the Dynes parameter describing the broadening of the bound state peaks. $\check K_i$ is kinetic term for nearest neighbor hopping
\begin{align}
\check K_i \check G_{\bm i \bm j} = \rho_\uparrow \sum \limits_{\bm a} \check G_{\bm i +\bm a -\bm a_y ,\bm j} + \rho_\downarrow \sum \limits_{\bm a} \check G_{\bm i +\bm a +\bm a_y ,\bm j}
\label{j}
\end{align}
where $\rho_{\uparrow,\downarrow}=(\rho_0\pm\rho_z)/2$ and $\bm a \in \{\pm\bm a_x, \pm\bm a_y, \pm\bm a_z \}$.
$\check U_{\bm i}$ is the potential of the impurity, which in the two-sublattice formalism takes the form:
\begin{align}
    \check U_{\bm i} = \check U_0 \delta_{\bm i,\bm i_{imp}}=U_0 \frac{(\rho_x+ \nu i\rho_y)}{2} \delta_{\bm i, \bm i_{imp}},
\end{align}
$\nu = \pm 1$ if the impurity is located at $A(B)$-site.

The standard $T$-matrix ansatz for the full Green's function takes the form:
\begin{align} 
    \check G_{\bm i \bm j}=\check G_{\bm i \bm j}^0+\sum\limits_{\bm m, \bm l}\check G_{\bm i \bm m}^0 \check T_{\bm m \bm l} \check G_{\bm l \bm j}^0,
    \label{green_T}
\end{align}
where $\check G_{\bm i,\bm j}^0$ is the Green's function of the homogeneous S/AF bilayer in the absence of impurity obeying 
\begin{eqnarray}
\check H_{\bm i}^0 \check G_{\bm i \bm j}^0 =  \delta_{\bm i \bm j}.
\label{eq_green_homogeneous}
\end{eqnarray}

Subtracting perturbed (\ref{Green_general}) and unperturbed (\ref{eq_green_homogeneous}) equations, and multiplying by $H_{\bm j}^0$ from the right, we obtain:
\begin{align}
    \check T_{\bm i\bm j}-\check  U_{\bm i} \sum\limits_{\bm m} \check G_{\bm i \bm l}^0 \check T_{\bm l\bm j} - \check U_{\bm i}\delta_{\bm i\bm j}=0
    \label{T_1}
\end{align}
Due to isotropy of impurity potential we can write
\begin{align}
    \check T_{\bm i \bm j}=\check T^0 \delta_{\bm i,\bm i_{imp}} \delta_{\bm j,\bm i_{imp}}.
    \label{T_1a}
\end{align}
Then from Eq.~(\ref{T_1}) it follows that
\begin{align}
    \check T^0=(1-\check U_0\check G_{\bm i_{imp},\bm i_{imp}}^0 )^{-1}\check U_0. 
    \label{T_2}
\end{align}
Energy of the bound states is determined by the poles of the $T$-matrix, that is the bound state energies are obtained from the following equation
\begin{align}
    \det (1-\check U_0 \check G_{\bm i_{imp},\bm i_{imp}}^0)=0
    \label{energy_bound_state}
\end{align}

The spin-resolved LDOS at A and B sites of the $\bm i$-th unit cell is determined via the imaginary part of the retarded Green's function
\begin{align}
    &N_{\uparrow,\downarrow}^{A,B}(\varepsilon,\bm i)= \nonumber \\ -&\frac{1}{\pi}{\rm Im}\left[{\rm Tr} [\frac{\check G_{\bm i \bm i} (\sigma_0+ s \sigma_z)(\tau_0+\tau_z)(\rho_x+\nu i\rho_y)}{8}]\right],
\end{align}
where $s = \pm 1$ for spin $\uparrow (\downarrow)$ and $\nu = \pm 1$ for $A(B)$ sublattices. The Green's function $\check G_{\bm i \bm i}$ is calculated according to Eqs.~(\ref{green_T}), (\ref{T_1a}), (\ref{T_2}). The homogeneous Green's function $\check G_{\bm i \bm j}^0$ can be calculated analytically:
\begin{align}
    \check G_{\bm i \bm j}^0=\int \frac{d^3\bm p}{(2\pi)^3} e^{\frac{\displaystyle -ip_y a_y \rho_z}{\displaystyle 2}}\check G^0(\bm p)e^{\frac{-\displaystyle ip_y a_y \rho_z}{\displaystyle 2}} e^{i \bm p (\bm i - \bm j)},
\end{align}
and $\check G^0(\bm p)$ should be found from unperturbed Gor'kov equation in the momentum representation:
\begin{align}
    (\varepsilon\tau_z+\mu+h\sigma_z\rho_z + i\tau_y \Delta)\rho_x \check G^0(\bm p) - \nonumber \\
    2t(\cos (p_x a)+\cos(p_y a))\check G^0(\bm p)=1 .
\end{align}
Then $\check G_{\bm i \bm j}^0$ is diagonal in spin space, that is $\check G_{\bm i \bm j}^0 = \check G_{\bm i \bm j,\uparrow}^0 (1+\sigma_z)/2 + \check G_{\bm i \bm j,\downarrow}^0 (1-\sigma_z)/2 $ and each of the components in spin space can be expanded over Pauli matrices in particle-hole and sublattice spaces:
\begin{align}
    G_{\bm i \bm i,\sigma}^0=G_{0x}^\sigma\tau_0\rho_x+G_{yx}^\sigma\tau_y\rho_x+G_{zx}^\sigma\tau_z\rho_x+ \nonumber \\      G_{0y}^\sigma\tau_0\rho_y+G_{yy}^\sigma\tau_y\rho_y+G_{zy}^\sigma\tau_z\rho_y
\end{align}
with
\begin{align}
    \begin{cases}
        G_{0x}^\sigma=\mu((-h^2+\Delta^2-\varepsilon^2+\mu^2)I_1-I_2) \\
        G_{yx}^\sigma=-i\Delta((-h^2+\Delta^2-\varepsilon^2+\mu^2)I_1+I_2) \\
        G_{zx}^\sigma=\varepsilon((-h^2-\Delta^2+\varepsilon^2-\mu^2)I_1-I_2) \\
        G_{0y}^\sigma=2i\sigma h\varepsilon \mu I_1 \\
        G_{yy}^\sigma=2 \sigma h\Delta \varepsilon I_1 \\
        G_{zy}^\sigma=i \sigma h ((h^2-\Delta^2-\varepsilon^2-\mu^2)I_1+I_2)
    \end{cases}
\end{align}
where $I_1=-\frac{2\sqrt{2}i\pi}{(\alpha_1+\alpha_2)\alpha_1\alpha_2}$, $I_2=\frac{\sqrt{2}i\pi}{\alpha_1+\alpha_2}$, $\alpha_{1,2}=\sqrt{-\beta_1\mp\sqrt{\beta_1^2-4\beta_0}}$, $\beta_0=h^4+(\Delta^2-\varepsilon^2+\mu^2)^2-2h^2(\Delta^2+\varepsilon^2+\mu^2)$, $\beta_1=2(\Delta^2-\varepsilon^2-\mu^2+h^2)$.

\section{Phase diagram}
\label{phase_diagram}

\begin{figure}[tb]
	\begin{center}
		\includegraphics[width=75mm]{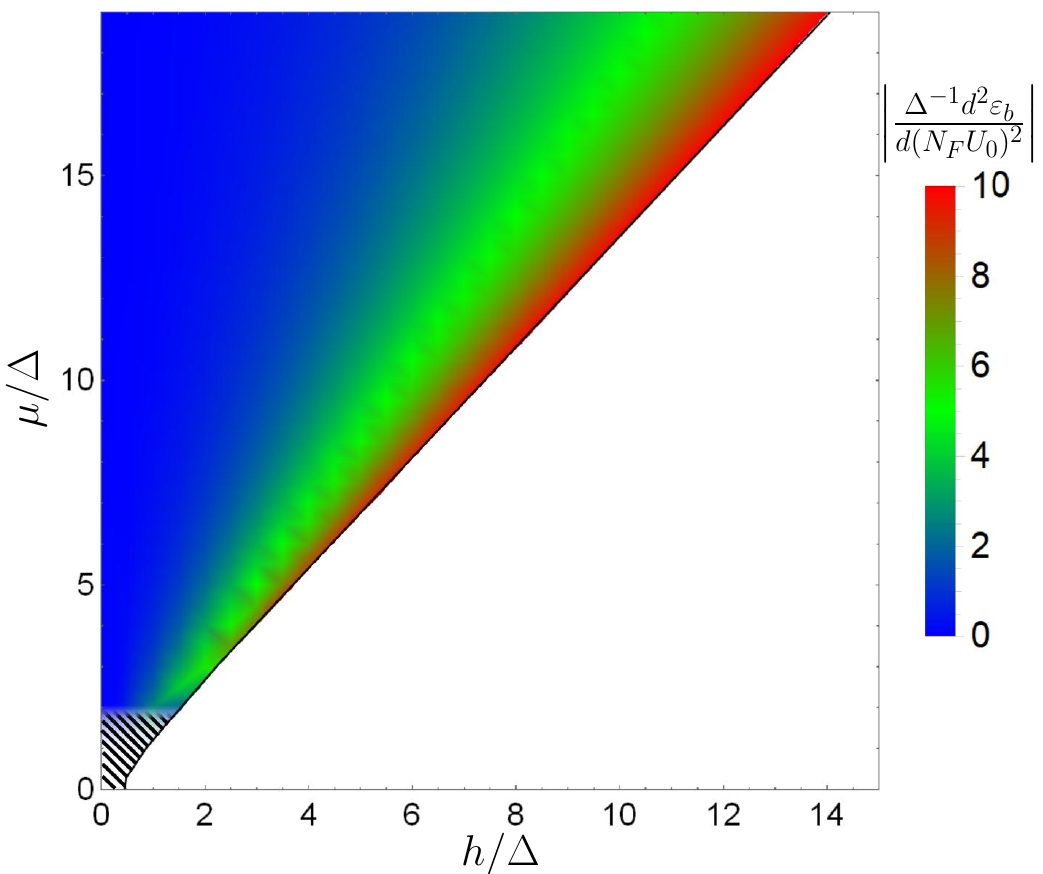}
\caption{$\bigl|d^2\varepsilon_b/dU_0^2 |_{U_0=0}\bigr|$ in the plane $(h,\mu)$. White region corresponds to fully suppressed superconductivity. }
 \label{fig:curvature}
	\end{center}
\end{figure}

\begin{figure}[tb]
	\begin{center}
		\includegraphics[width=75mm]{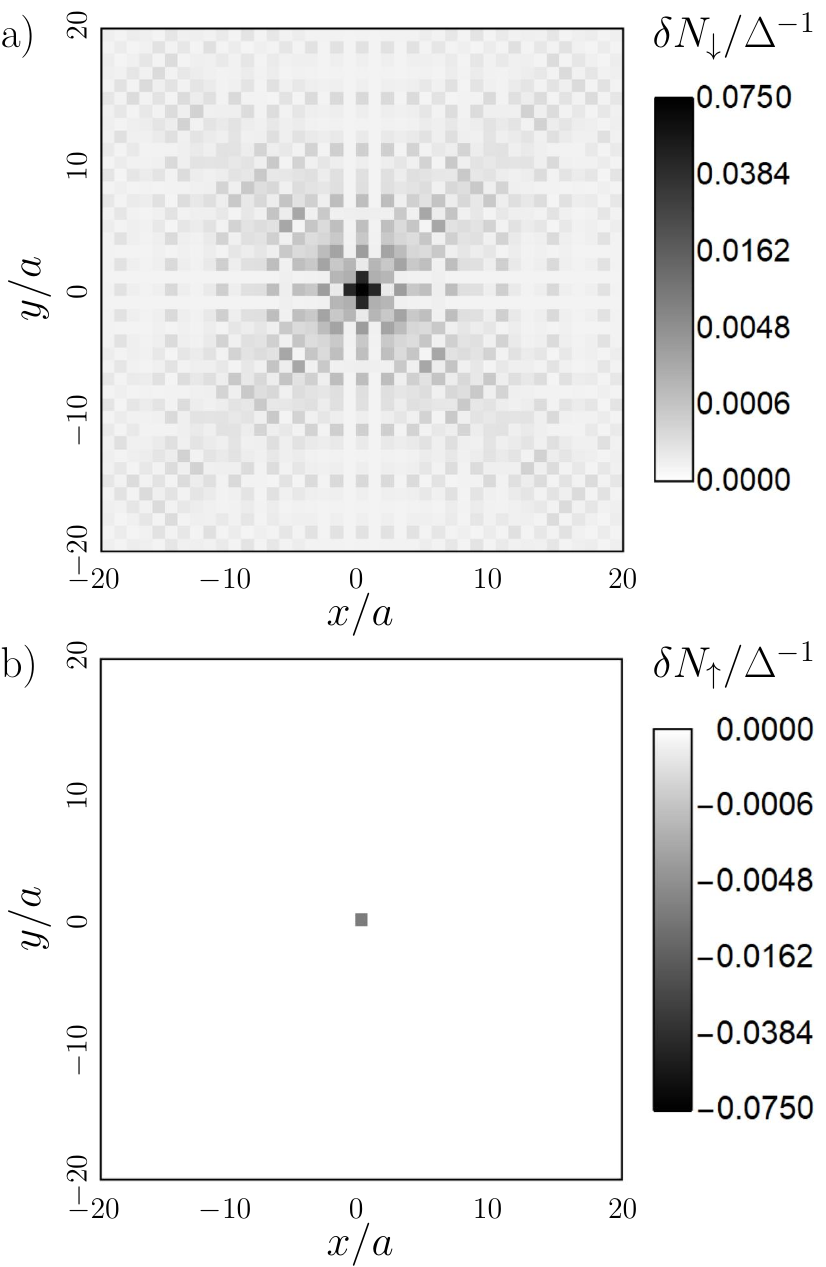}
\caption{Perturbation $\delta N_\sigma = N_\sigma - N_{0,\sigma}$ of the LDOS by the impurity. Here $N_{0,\sigma}$ is the LDOS at the same site in the absence of the impurity. (a) Spin-down local perturbation of the LDOS $\delta N_\downarrow$ and (b) spin-up perturbation $\delta N_\uparrow$. $\mu=20\Delta$, $h=15\Delta$, $U_0 = 10\Delta$.}
 \label{fig:spin}
	\end{center}
\end{figure}

\begin{figure}[tb]
	\begin{center}
		\includegraphics[width=75mm]{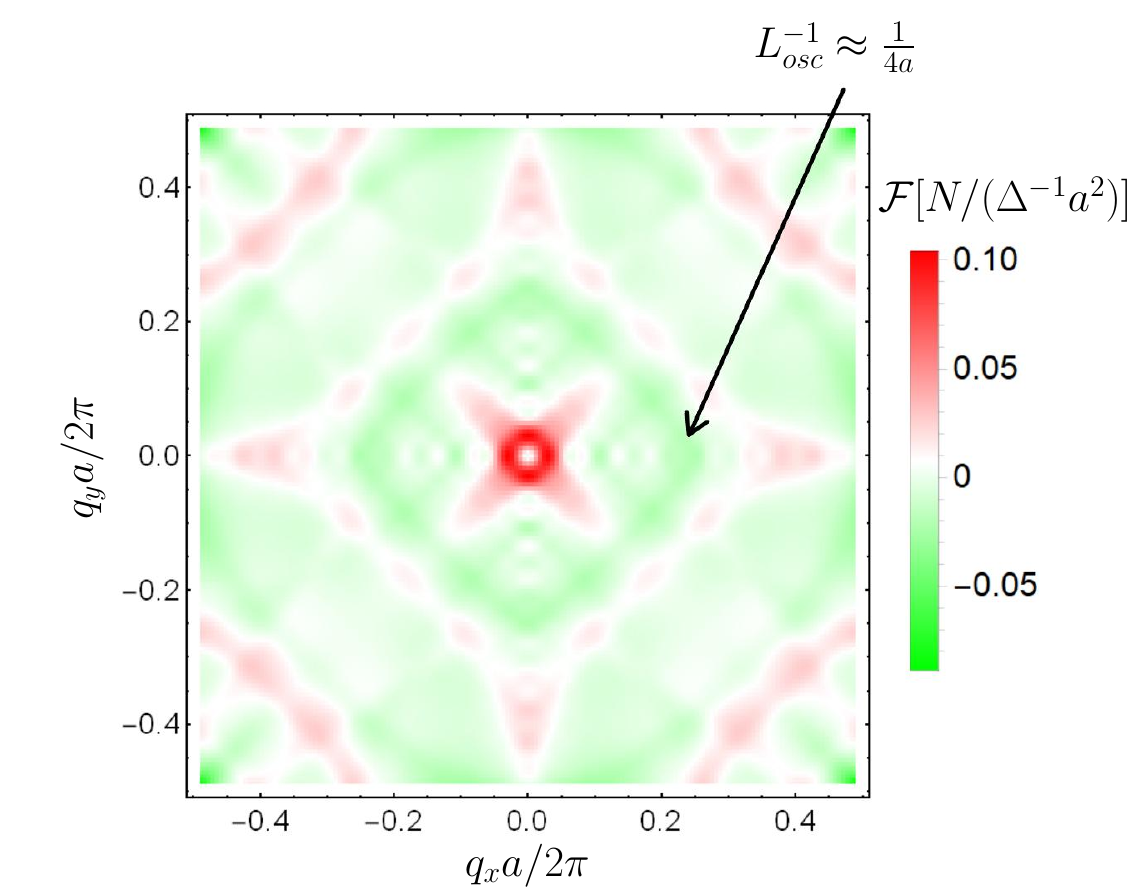}
\caption{Fourier transform of the LDOS presented in Fig.~\ref{fig:LDOS_spatial}.  $\varepsilon = -\varepsilon_b$. The impurity is at $A$-site. $\mu=20\Delta$, $h=15\Delta$, $U_0 = 10\Delta$.}
 \label{fig:fourier}
	\end{center}
\end{figure}

In order to provide more information on the phase diagram  of the bound state existence in Fig.~\ref{fig:curvature} we present the curvature $\bigl|d^2\varepsilon_b/dU_0^2 |_{U_0=0}\bigr|$ in the plane $(h,\mu)$. If this quantity is more than zero, it indicates the existence of the bound states for a given set $(h, \mu)$ at an arbitrary value of $U_0$. This figure shows the same trends as Fig.~\ref{fig:phase_diagram}, thus indicating that the regions of existence/absence of the bound states do not depend qualitatively on the particular value of $U_0$. The shaded area corresponds to the region of small $\mu$, where the bound states do not exist at all values of $U_0$ and, therefore, it is not possible to determine the curvature.

\section{Spin structure of the bound state LDOS}

\label{spin_structure}

\begin{figure}[tb]
	\begin{center}
		\includegraphics[width=75mm]{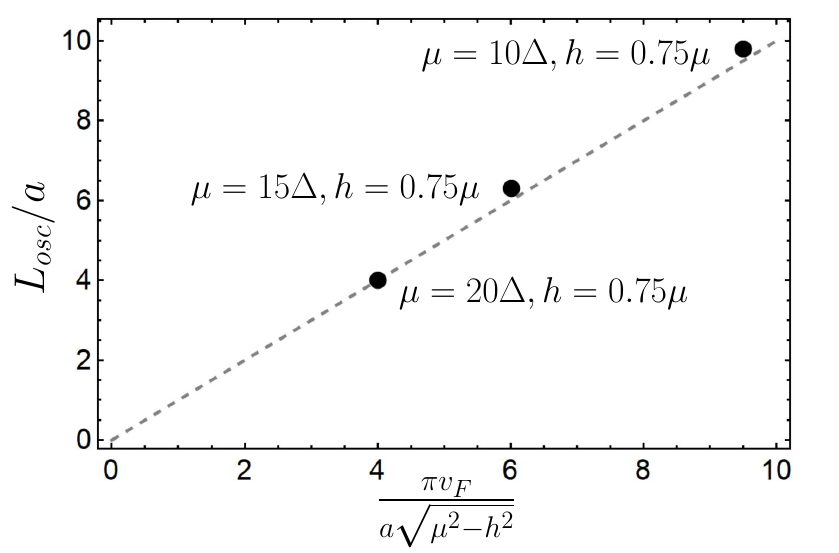}
\caption{$L_{osc}$ extracted from the LDOS for different points in the $(h,\mu)$-space, denoted by the black circles. Dashed line represents the formula $L_{osc} = \pi v_F/\sqrt{\mu^2-h^2}$.}
 \label{fig:period}
	\end{center}
\end{figure}

\begin{figure}[tb]
	\begin{center}
		\includegraphics[width=75mm]{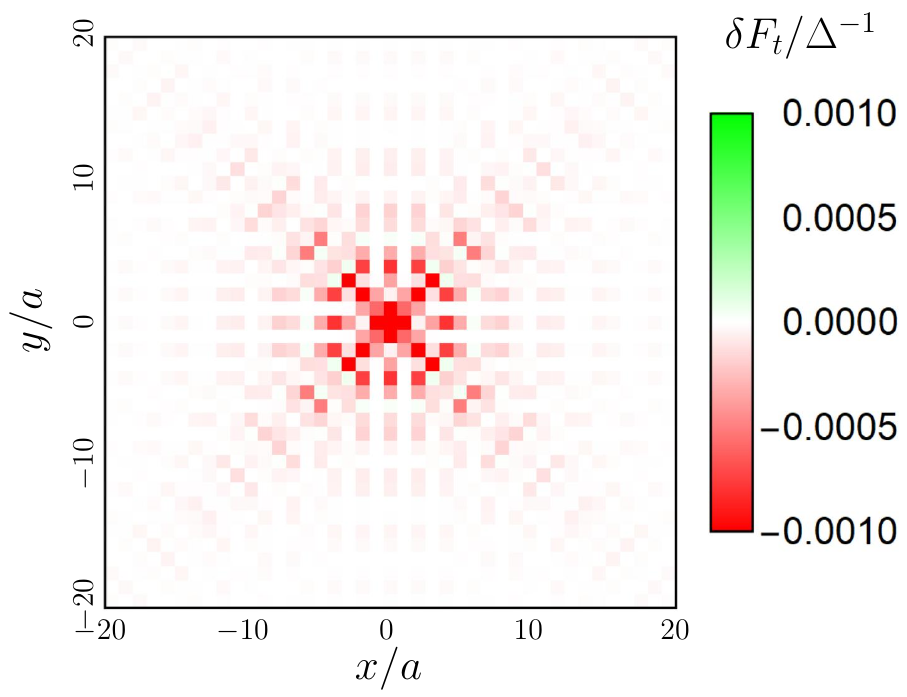}
\caption{Impurity-induced perturbation $\delta F_t$ of the ideal N\'eel triplet correlation structure at  $\varepsilon = -\varepsilon_b$. The impurity is at $A$-site. $\mu=20\Delta$, $h=15\Delta$, $U_0 = 10\Delta$.}
 \label{fig:triplets}
	\end{center}
\end{figure}

In order to prove that the Andreev bound states generated by a nonmagnetic impurity are spin-resolved, in Fig.~\ref{fig:spin} we plot the spin-resolved LDOS at $\varepsilon = - \varepsilon_b$. Only the perturbation $\delta N_\sigma$ of the LDOS by the impurity is shown. Here $\delta N_\sigma = N_\sigma - N_{0,\sigma}$, where $N_{0,\sigma}$ is the LDOS at the same site in the absence of the impurity. It manifests a perfect staggered order. In Fig.~\ref{fig:spin} the nonmagnetic impurity is located at $A$-site. It is seen that in this case only the spin-down LDOS $N_\downarrow$ is perturbed by the impurity for the lower bound state $-\varepsilon_b$, except for the impurity site. Directly at the impurity site we see very local perturbation of the LDOS in the spin-up subband, which results from the fact that the nonmagnetic impurity by itself works as a on-site perturbation of the chemical potential.

\section{Spatial oscillations of the LDOS}
\label{appendix_spatial}

Here we provide more detailed data that prove that the reason for the appearance of  the additional period of the LDOS oscillations is the finite-momentum N\'eel triplet pairing. In Fig.~\ref{fig:fourier} we demonstrate the Fourier transform of the LDOS data presented in Fig.~\ref{fig:LDOS_spatial}. The peak corresponding to the oscillation period $L_{osc} \approx 4a$ is clearly seen as a green ring curve. Also in Fig.~\ref{fig:LDOS_spatial} we see two additional specific features. The largest green ring curve approximately having radius $qa \approx \pi$ represents the N\'eel staggered order of the LDOS. The central red ring curve of the smallest radius originates from the overall decay of the impurity-induced LDOS at the length scale $\xi$. The four-fold-symmetric anisotropy of the image is due to the fact that we consider square lattice.

In Fig.~\ref{fig:period} we additionally demonstrate the oscillation period $L_{osc}$ extracted from the LDOS for different points in the $(h,\mu)$-space. Dashed line represents the formula $L_{osc} = \pi v_F/\sqrt{\mu^2-h^2}$. It is seen that the data are in excellent agreement with this dependence. Moreover, as it was already indicated in Sec.~\ref{spatial}, the period increases as $\mu$ decreases. This suggests that this periodic pattern cannot be ascribed to the Friedel oscillations because in the framework of the considered  tight-binding model on a square lattice the period of Friedel oscillations does not manifest such a monotonic dependence on $\mu$.

Also in Fig.~\ref{fig:triplets} we demonstrate the perturbation in the spatial structure of the N\'eel-type triplet correlations induced by the impurity. The perturbation is defined as $\delta F_t(\varepsilon) = F_t(\varepsilon) - F_t^0(\varepsilon)$, where $F_t^{A,B}(\varepsilon) = (1/8){\rm Tr}\left[ \check G_{\bm i \bm i}(\varepsilon)(\tau_x - i \tau_y)\sigma_z(\rho_x+i\nu \rho_y)  \right]$ is the anomalous component of the retarded Green's function and $F_t^{0 A,B}(\varepsilon) = (1/8){\rm Tr}\left[ \check G_{\bm i \bm i}^{0}(\varepsilon)(\tau_x - i \tau_y)\sigma_z(\rho_x+i\nu \rho_y)  \right]$ is the anomalous component of the homogeneous Green's function in the absence of the impurity. The Green's function $\delta F_t$ in Fig.~\ref{fig:triplets} is a sum of the both sublattices. The anomalous Green's function is plotted at $\varepsilon = -\varepsilon_b$. The rings corresponding to the oscillations of the amplitude of the N\'eel triplet pairs with the period $L_{osc} \approx 4a$ are seen in this figure.

\bibliography{impurities}

\end{document}